\documentclass{article}
\usepackage{graphicx}        

\usepackage{graphicx}
\usepackage{caption} 
\usepackage{subcaption}

\usepackage{mathtools}
\usepackage{bm}
\usepackage{authblk}

\newcommand{\vel}{\mathbf{u}}
\newcommand{\divtau}{\nabla \cdot \boldsymbol{\tau}}
\newcommand{\rfv}{\bm{t}}
\newcommand{\arfv}{\tilde{\bm{t}}}
\newcommand{\orfv}{\bm{t}^\perp}
\newcommand{\aorfv}{\widetilde{\bm{t}^\perp}}
\newcommand{\tr}{\textnormal{tr}}

\newcommand{\RANS}{\textnormal{RANS}}
\newcommand{\DNS}{\textnormal{DNS}}
\newcommand{\NN}{\textnormal{NN}}
\newcommand{\VBNN}{\textnormal{VBNN}}

\begin{document}

\title{Improving the Vector Basis Neural Network for RANS Equations Using Separate Trainings}

\author[$\star$]{D. Oberto}
\affil[$\star$]{Dipartimento di Scienze Matematiche, Politecnico di Torino, Corso Duca degli Abruzzi 24, 10129 Torino, Italy. \hspace{2cm} Email: davide.oberto@polito.it}
%
%
\maketitle

\abstract{We present a new data-driven turbulence model for Reynolds-averaged Navier-Stokes equations called $\nu_t$-Vector Basis Neural Network. This new model, grounded on the already existing Vector Basis Neural Network, predicts separately the turbulent viscosity $\nu_t$ and the contribution of the Reynolds force vector that is not already accounted in $\nu_t$. Numerical experiments on the flow in a Square Duct show the better accuracy of the new model compared to the reference one.}

\section{Introduction}
\label{sec:Intro}
Directly solving Navier-Stokes (DNS) equations is computationally unaffordable for the majority of turbulent industrial-relevant flows. Indeed, solving the Reynolds-averaged Navier-Stokes (RANS) equations is the common choice for such flows. RANS equations are obtained by formally applying an average operator to the classic Navier-Stokes equations and solving for the averaged fields \cite{Pope}. This procedure leads to deal with significantly coarser computational space-time meshes and, consequently, lower computational cost.

However, RANS equations have more unknowns than equations and need to be closed using \emph{turbulence models}. These generally consists in additional heuristically-derived partial differential equations. Despite having decades of history, RANS equations coupled with classic turbulence models can be particularly inaccurate for some classes of flows. 

This issue has driven researchers to exploit Scientific Machine Learning techniques to take advantage of the available DNS simulations to train data-driven models aiming to overtake  traditional turbulence models when the latter fail \cite{Review}. In this context, data-driven models should satisfy the same physical properties that classic turbulence models guarantee, namely Galilean invariance and frame-reference independence \cite{Ling_rotations,Ling}.

Different models have been proposed in the recent years. They usually differ from each other by the chosen field to predict through machine learning, the strategies to guarantee physical properties to the model and the resulting momentum equation of the RANS system to solve. In this work, we propose a new model called $\nu_t$-Vector Basis Neural Network ($\nu_t$-VBNN in the following) that is grounded on the Vector Basis Neural Network (VBNN) proposed in \cite{nostro}, but that differs on the selected turbulent fields to predict and, consequently, on the resulting momentum equation. In particular, the $\nu_t$-VBNN model predicts separately through neural networks the turbulent viscosity $\nu_t$ and the remaining turbulent contribution as a vector instead of predicting a unique vector (the Reynolds force vector) as for the VBNN setting.

This paper is organized as follows: in Section \ref{sec:RANS} we give an overview on RANS equations and on the necessity of turbulence modeling; in Section \ref{sec:data-driven RANS} we generally describe the role of Machine Learning in RANS turbulence models and we describe in detail the VBNN defined in \cite{nostro} and the new $\nu_t$-VBNN model; finally the accuracy of the two models are compared in Section \ref{sec:results} for the flow in a square duct showing that the new $\nu_t$-VBNN model outperforms the VBNN one \cite{nostro}.

\section{Reynolds-Averaged Navier-Stokes Equations} \label{sec:RANS}

For incompressible steady flows, RANS equations read
\begin{equation} \label{eq:RANS}
\begin{dcases}
&\nabla \cdot \vel=0 \\
&\vel \cdot \nabla\vel - 2 \nabla \cdot [\nu \bm{S}] = - \nabla p - \divtau,
\end{dcases}
\end{equation}
where the first equation is obtained from the mass balance law while the second one from the momentum balance law. In \eqref{eq:RANS}, $\nu$ is the kinematic viscosity of the fluid, $\vel$ is the averaged velocity field,  $p$ is the averaged pressure field divided by the density of the fluid and $\bm{S}$ is the symmetric part of the gradient of $\vel$. Finally, $\boldsymbol{\tau}$ is an unknown second order symmetric tensor called \emph{Reynolds stress tensor} (RST) that needs to be modeled to close RANS equations and whose components are associated to the turbulent fluctuations of the velocity field. 

Let us define the \emph{Reynolds force vector} (RVF) as $\bm{t} \coloneqq \divtau$. Looking to RANS equations \eqref{eq:RANS}, the RFV is the only term associated to not-averaged fields. Consequently, we can interpret the RFV as the turbulence effects on the averaged fields.

Classically, the RANS equations are closed by adding additional partial differential equations with unknowns either the RST components or some turbulent fields associated to it \cite{Pope}. The well-known $k-\varepsilon$ model is an example of the second class in which two transport equations are written to obtain the trace of the RST called \emph{turbulent kinetic energy} $k \coloneqq \tr({\boldsymbol{\tau}})$ and its dissipation rate $\varepsilon$. Once we have these two fields, the RST is obtained using the so-called Boussinesq's assumption:

\begin{equation} \label{eq:linear_closures}
\boldsymbol{\tau} = \frac{2}{3} k \mathbf{I} - 2 \nu_t \mathbf{S},
\end{equation} 

where $\mathbf{I}$ is the identity tensor and $\nu_t = 0.09 k^2/\varepsilon$ is the \emph{turbulent viscosity}. By inserting \eqref{eq:linear_closures} into \eqref{eq:RANS}, we obtain a new momentum equation

\begin{equation} \label{eq:nut_momentum}
\vel \cdot \nabla\vel - 2 \nabla \cdot [(\nu + \nu_t) \bm{S}] = - \nabla \hat{p},
\end{equation} 
with $\hat{p} = p + 2/3 k$ and the system is now closed.

\section{Data-driven RANS Closures}  \label{sec:data-driven RANS}
%
Data-driven RANS models exploit data coming from high-fidelity DNS simulations to learn a mapping between averaged fields, available through RANS, and turbulent fluctuating fields, such as $\tau$ or $\bm{t}$ \cite{primo_Cruz}. 

Once $\tau$ or $\bm{t}$ are obtained, they have to be inserted into the momentum equation in \eqref{eq:RANS}. The most straightforward approach of treating them as source term into the momentum equation leads to inaccurate solutions even if the Machine Learning approach is well-trained and accurate. This bad conditioning phenomenon has been reported, for instance, in \cite{cond_Wu, cond_Cruz, nostro}. 

To alleviate this issues, the most common approach in literature consists of splitting either $\tau$ or $\bm{t}$ into two terms, one that can be written in terms of a viscous contribution and the remaining part. The former is then treated implicitly into the diffusive term of the momentum equation while the latter is still treated explicitly as forcing term. \\

In the next sections, we firstly present the VBNN \cite{nostro}, describing the implicit treatment of the diffusive term and the physical properties it preserves. Successively, we discuss a new method called $\nu_t$-VBNN. The latter is grounded on the same rational of the former with a significant change on the implicit treatment of the viscous term and that guarantees the same physical properties of the VBNN model.

\subsection{The Vector Basis Neural Network} 
Let us define the \emph{dimensionless Reynolds force vector} $\arfv \coloneqq k^{1/2}/\varepsilon \ \rfv$ and let us make the following assumption:
\begin{equation} \label{eq:divtau_const_hp}
\arfv = \mathbf{f}(\bm{s},\bm{w},\widetilde{\nabla \cdot \bm{S}},\widetilde{\nabla k}, Re_d),
\end{equation}
where $\bm{s} \coloneqq k / \varepsilon \ \bm{S}$, $\bm{w} \coloneqq k / \varepsilon \ \bm{W}$ being $\bm{W}$ the antisymmetric part of the mean velocity gradient, $\widetilde{\nabla \cdot \bm{S}} = k^{5/2} / \varepsilon^2 \  \nabla \cdot \bm{S}$ and $\widetilde{\nabla k} = k^{1/2} / \varepsilon \ \nabla k$. All these quantities are dimensionless. Finally $Re_d = \min(\frac{\sqrt{k} d}{50 \nu},2)$ is a scalar field called wall Reynolds number that depends on the distance $d$ from wall boundaries.

Using \eqref{eq:divtau_const_hp} and some representation theorems \cite{nostro}, we can write
\begin{equation} \label{eq:lin_comb}
\arfv = \sum_{k=1}^{12}{c_k (\lambda_1,\dots,\lambda_{27}) \ \bm{v}_k},
\end{equation} 
where $\{\lambda_j\}_{j=1}^{27}$ are known scalar fields, usually referred as \emph{invariants}, and $\{\bm{v}_k\}_{k=1}^{12}$ are known vector fields. To have a complete list of these fields, we refer to \cite{nostro}.

The model in \cite{nostro} uses a neural network, called Vector Basis Neural Network, that is trained to predict the coefficients of the linear combination in \eqref{eq:lin_comb} using RANS inputs, aiming to obtain a RFV close to the DNS one. In particular, the optimization process used to tune the weights of the neural network tries to minimize
\begin{equation} \label{eq:to_minimize}
\| \ \arfv^{\ \textnormal{DNS}} - \sum_{k=1}^{12}{c^\NN_k (\lambda^\RANS_1,\dots,\lambda^\RANS_{27}) \ \bm{v}^\RANS_k} \ \|_2.
\end{equation}
The fields $\lambda^\RANS_j$, $j=1,\dots,27$, and $\bm{v}^\RANS_k$, $k = 1,\dots,12$, come from RANS simulations and $c^\NN_k$, $k = 1,\dots,N_c$, are the VBNN outputs. In \eqref{eq:to_minimize}, we define $\arfv^{\ \textnormal{DNS}}=(k^{1/2})^\RANS / \varepsilon^\RANS \ \rfv^{\ \textnormal{DNS}}$. \\

By exploiting that $\bm{v}_1 = \widetilde{\nabla \cdot \bm{S}}$, we can write 
\begin{equation} \label{eq:turb_like}
\rfv = \nu_{tl} \ \nabla \cdot \mathbf{S} + \frac{\varepsilon}{k^{1/2}} \sum_{k=2}^{12}{c_k \ \bm{v}_k},
\end{equation}
where $\nu_{tl} = k^2 / \varepsilon \ c_1$ is called \emph{turbulent-like viscosity}. In \eqref{eq:turb_like}, we dot not write explicitly the dependence of the coefficients from the invariances for the sake of conciseness. 

The final momentum equation obtained using \eqref{eq:turb_like} into RANS equations \eqref{eq:RANS} reads
\begin{equation} \label{eq:mom_nutl}
\vel \cdot \nabla\vel - 2 (\nu + \nu_{tl}) \nabla \cdot \bm{S} = - \nabla \hat{p} - \bm{t}^\dag
\end{equation}
with $\bm{t}^\dag \coloneq \varepsilon / k^{1/2} \ \sum_{k=2}^{12}{c_k \ \bm{v}_k}$. \\

Finally, as observed in \cite{nostro}, this setting enforces both Galilean invariance and frame-reference independence. In particular, the former means that the VBNN is invariant with respect frames that differ by constant velocities and it is satisfied because all inputs and the vector basis are Galilean invariant (they depend on the gradient of $\vel$ and not on $\vel$ itself). The latter states that, if $\arfv$ is the output of the VBNN in a frame, and, if we describe our system in a new frame obtained by a rotation matrix $\bm{Q}$ of the former, then the new output of the VBNN must be $\arfv^{\ \bm{Q}} \coloneqq \bm{Q} \arfv$. This is true:

\begin{equation} \label{eq:rotation}
\arfv^{\ \bm{Q}} = \sum_{k=1}^{12}{c_k (\lambda_1^{\ \bm{Q}},\dots,\lambda^{\ \bm{Q}}_{27}) \ \bm{v}^{\ \bm{Q}}_k} = \sum_{k=1}^{12}{c_k (\lambda_1,\dots,\lambda_{27}) \ \bm{Q} \bm{v}_k} = \bm{Q} \arfv,
\end{equation}
where we have exploited $\lambda^{\ \bm{Q}}_i = \lambda_i$ and $\bm{v}^{\ \bm{Q}}_k = \bm{Q} \bm{v}_k$ being the former scalar fields and the latter vector fields.

\subsection{The Turbulent-Viscosity Vector Basis Neural Networks}
Following \cite{cond_Wu, cond_Cruz}, we can obtain the \emph{turbulent viscosity} as 
\begin{equation} \label{eq:nut}
\nu_t = - \frac{1}{2} \frac{\bm{\tau : S} }{\bm{S : S}},
\end{equation}
where $\bm{:}$ stands for the tensor scalar product, and define the orthogonal component of the RST as $\bm{\tau}^\perp = \bm{\tau} + 2 \nu_t \bm{S}$. For the sake of brevity, we will denote $\bm{t}^\perp \coloneqq \nabla \cdot \bm{\tau}^\perp$.

The $\nu_t$-VBNN model uses two different neural networks: one to obtain $\nu_t$ and one to predict $\bm{t}^\perp$. The former ($\NN_{\nu_t}$ in the following) is a simple feed forward neural network with the invariants in \eqref{eq:lin_comb} as inputs and a unique output. The latter ($\VBNN_{\orfv}$ in the following) has the same architecture of the VBNN one. The only difference consists in the different target vector field that reflects in the different coefficients of the linear expansion in \eqref{eq:lin_comb}. Indeed, the two neural networks are trained separately by minimizing
\begin{equation} \label{eq:to_minimize_2}
\begin{split}
& \| \nu_t^\DNS - \nu_t^\NN(\lambda^\RANS_1,\dots,\lambda^\RANS_{27}) \|, \\
\| \ \aorfv^{\ \textnormal{DNS}} &  - \sum_{k=1}^{12}{c^{\perp,\NN}_k (\lambda^\RANS_1,\dots,\lambda^\RANS_{27}) \ \bm{v}^\RANS_k} \ \|_2,
\end{split}
\end{equation}
respectively, where $\aorfv = k^{1/2}/\varepsilon \ \orfv$. 

The momentum equation associated to the $\nu_t$-VBNN method reads
\begin{equation} \label{eq:mom_nut}
\vel \cdot \nabla\vel - 2 \nabla \cdot[(\nu + \nu_{t}) \bm{S}] = - \nabla \hat{p} - \orfv.
\end{equation}
This equation differs from the VBNN one \eqref{eq:mom_nutl} by the diffusion term and the forcing term. Additionally, equation \eqref{eq:mom_nut} can be interpreted as a generalization of equation \eqref{eq:nut_momentum} with the addition of a forcing term.

Finally, we observe that the $\nu_t$-VBNN approach satisfies Galilean invariance and frame-reference independence using the same arguments of the VBNN case. Specifically, the inputs and the vector basis used for $\VBNN_{\orfv}$ are Galilean invariants. Additionally, $\NN_{\nu_t}$, that predicts a scalar field, does not change with rotations of the frame while the $\VBNN_{\orfv}$ correctly rotates the output field as for the VBNN case.

\section{Numerical Results} \label{sec:results}
\subsection{Duct Flow Test Case}
\begin{figure}[h]
\centering
\includegraphics[width=0.65\textwidth] {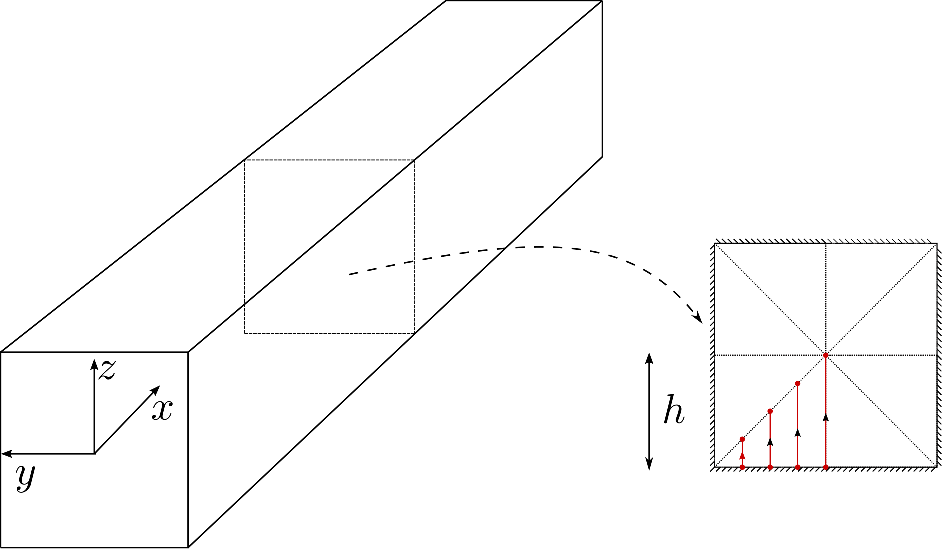}
    \caption{Square duct domain and square $yz$-section.}
\label{fig:square_duct}
\end{figure}

We test the VBNN and the $\nu_t$-VBNN for the flow in a square duct. Figure \ref{fig:square_duct} shows the square duct domain and a $yz$-square section. The height of the square section is $2h$, the duct length is $L = 4 h \pi$. Periodic boundary conditions are set at the inlet ($x=0$) and outlet ($x=L$) faces and no-slip boundary conditions are set on the remaining boundaries.

This flow is a classic benchmark for data-driven turbulence models for two main reasons: i) availability of DNS results. Indeed, we exploit DNS results from \cite{Pinelli} at various bulk Reynolds numbers $Re_b = U_b h /\nu$, where $U_b$ is the average of $u_x$ across a square section; ii) difficulty of classic RANS turbulence models to describe the secondary motion associated to non-zero $u_y$ and $u_z$ components.

\subsection{VBNN and $\nu_t$-VBNN Trainings}
We generate a dataset by performing RANS simulations corresponding to available DNS ones at $Re_b = 2000, 2200, 2400, 2600, 2900, 3200, 3500$. Data coming from the $Re_b = 2900$ case are kept out from the training process and used for testing purposes only. We use a classic $k-\varepsilon$ turbulence model based on assumption \eqref{eq:linear_closures} as classic RANS model to improve. We will refer to it as Baseline model. All simulations are carried out using the finite volume-based open source OpenFoam code.

To obtain a one-to-one correspondence between DNS and RANS fields to evaluate either \eqref{eq:to_minimize} or \eqref{eq:to_minimize_2}, we interpolate DNS data on the RANS mesh, obtaining around $7.7 \cdot 10^5$ training cells.

Both VBNN and $\nu_t$-VBNN have 8 hidden layers of 30 neurons each. They are constrained to have 27 input neurons (as many as the invariants in \eqref{eq:lin_comb}). VBNN and $\VBNN_{\orfv}$ have 12 output neurons (as many as the coefficients in \eqref{eq:lin_comb}) while $\NN_{\nu_t}$ has one output only. All neural networks have exponential linear unit activation functions and are trained using an Adam optimization scheme using a learning rate exponentially decreasing during the training.

\subsection{Velocity Field Analysis}
Once either VBNN or $\nu_t$-VBNN are trained, one can run a new Baseline RANS simulation, compute the invariants and the vector basis, predict the turbulent fields, insert them into the momentum equations \eqref{eq:mom_nutl} or \eqref{eq:mom_nut} and run a new simulation. In both cases, the new velocity field should be closer to the DNS one compared to the Baseline case. For the sake of simplicity, we will refer to the simulations using the VBNN and $\nu_t$-VBNN models with the same names of the models themselves. As mentioned earlier, we compare the results using the Baseline, the VBNN and the $\nu_t$-VBNN models with the DNS results for the case of $Re_b=2900$ that is not used during the training.

\begin{figure}[h]
\centering
\begin{subfigure}{0.45\textwidth}
    \includegraphics[width=\textwidth] {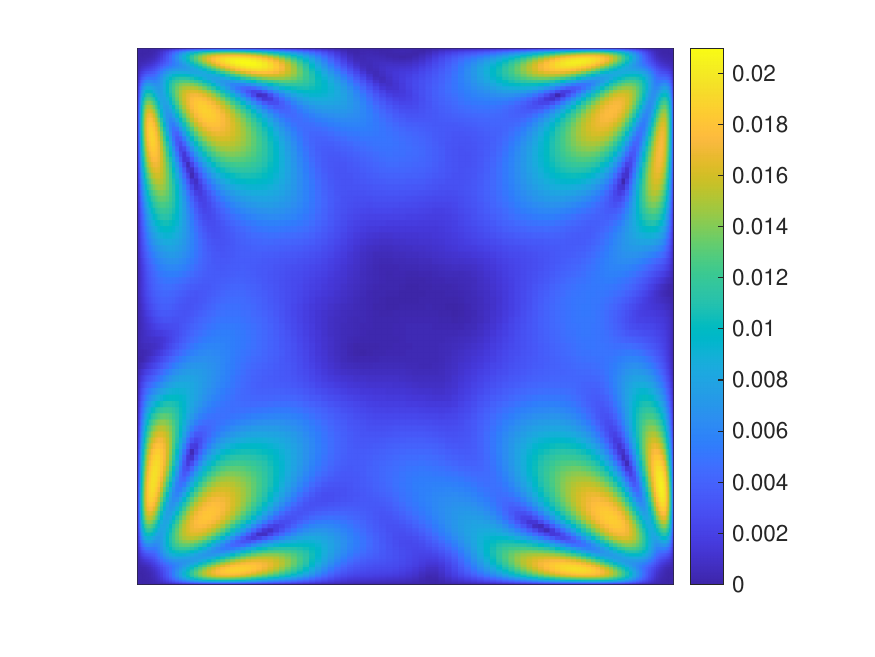}
    \caption{DNS}
\end{subfigure}\,
    \begin{subfigure}{0.45\textwidth}
    \includegraphics[width=\textwidth] {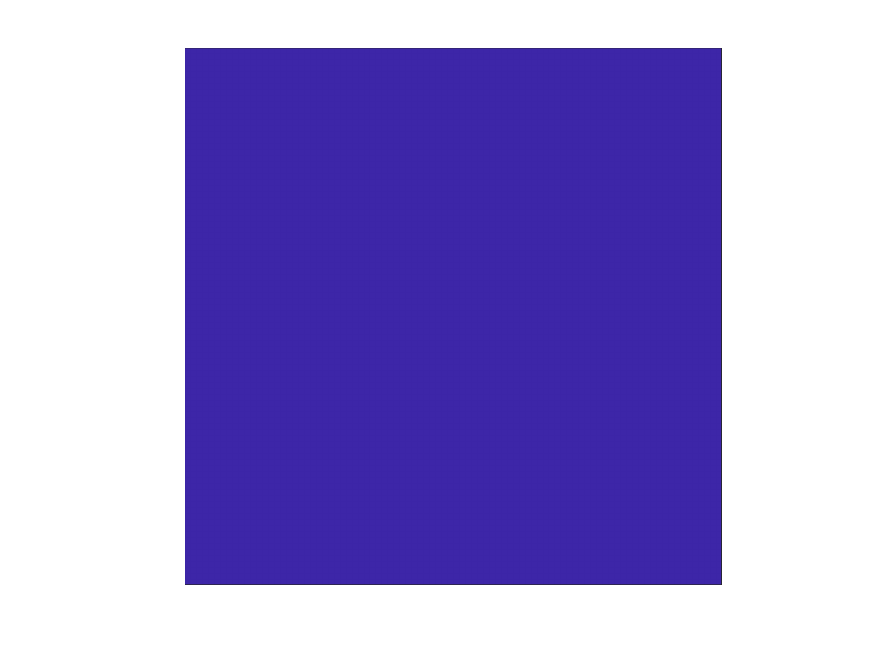}
    \caption{Baseline}
\end{subfigure}\,
\begin{subfigure}{0.45\textwidth}
    \includegraphics[width=\textwidth] {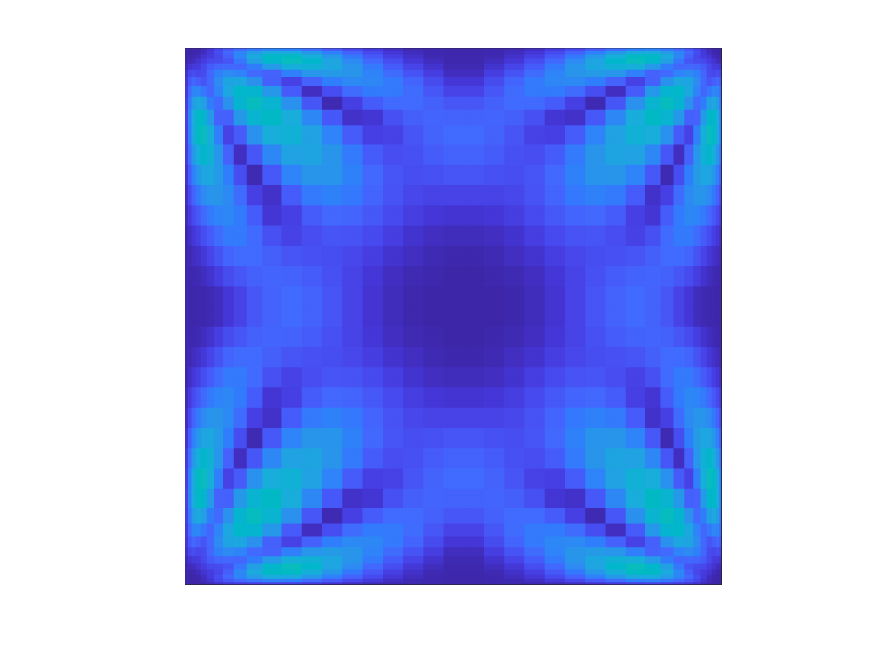}
    \caption{VBNN}
\end{subfigure}\,
\begin{subfigure}{0.45\textwidth}
    \includegraphics[width=\textwidth] {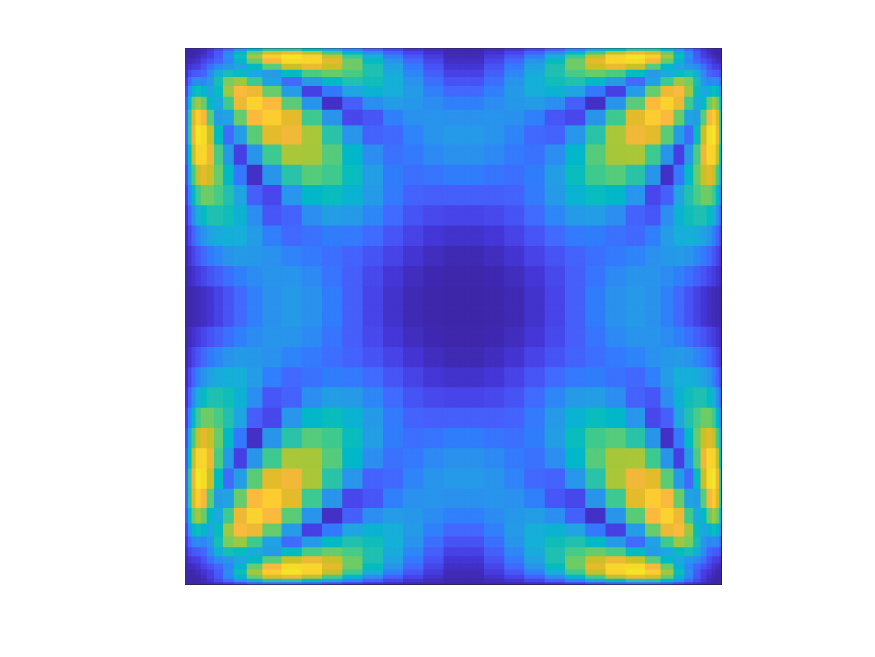}
    \caption{$\nu_t$-VBNN}
\end{subfigure}
\caption{Intensity of the secondary motion $\| (u_y, u_z)^T \|_2 / U_b$.}
\label{fig:secmot}
\end{figure}

Figure \ref{fig:secmot} compares the intensity of the secondary, defined as $\| (u_y, u_z)^T \|_2 / U_b$, in the four simulations. In addition, maxima values of the secondary motion and their ratio with the DNS reference one are listed in Table \ref{Tab:second_motion}.  It immediately appears that, as expected, the Baseline model does not predict any secondary motion. On the other hand, there is secondary motion for both VBNN and $\nu_t$-VBNN. However, the VBNN result drastically underestimates it while the $\nu_t$-VBNN case is significantly more accurate.

\begin{table}[h]
\caption{Maxima of the secondary motion norm and corresponding amplification factor}
\begin{center}
\begin{tabular}{*{3}{c}} 
\hline
model & $\max(\vert \vert(u_y,u_z)^T\vert \vert_2)/u_b$ & $\frac{\max(\vert \vert(u_y,u_z)^T\vert \vert_2)}{\max(\vert \vert(u_y^\DNS,u_z^\DNS)^T\vert \vert_2)}$ \\
\hline
DNS & 2.01 e-2 & 1 \\
VBNN & 1.03 e-2 & 0.489 \\
$\nu_t$-VBNN & 1.98 e-2 & 0.945 \\
Baseline & 4.31 e-14 & 0.000 \\
\hline
\end{tabular}
\label{Tab:second_motion}
\end{center}
\end{table}

\begin{figure}[h]
\centering
\begin{subfigure}{0.45\textwidth}
    \includegraphics[width=\textwidth] {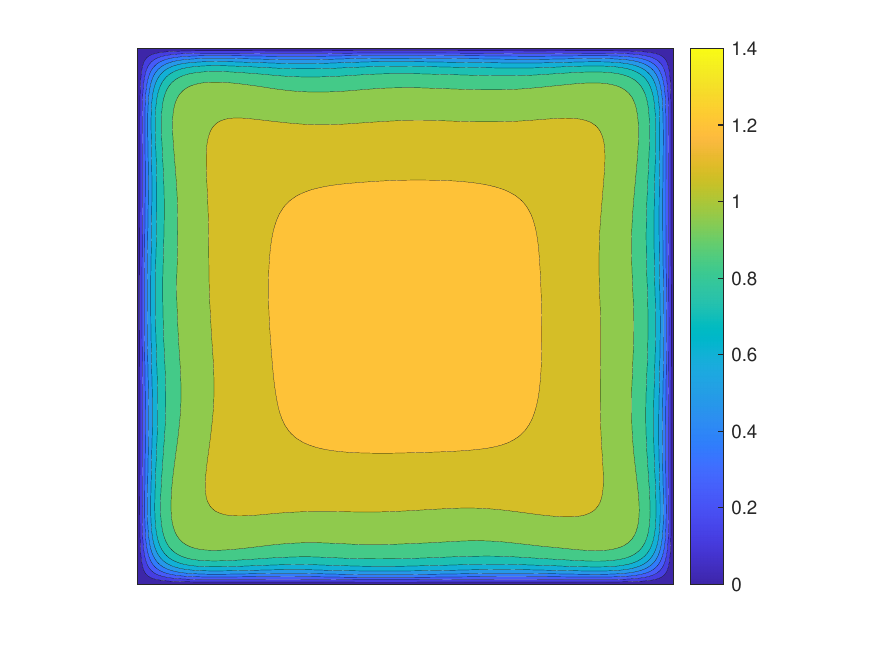}
    \caption{DNS}
\end{subfigure}\,
    \begin{subfigure}{0.45\textwidth}
    \includegraphics[width=\textwidth] {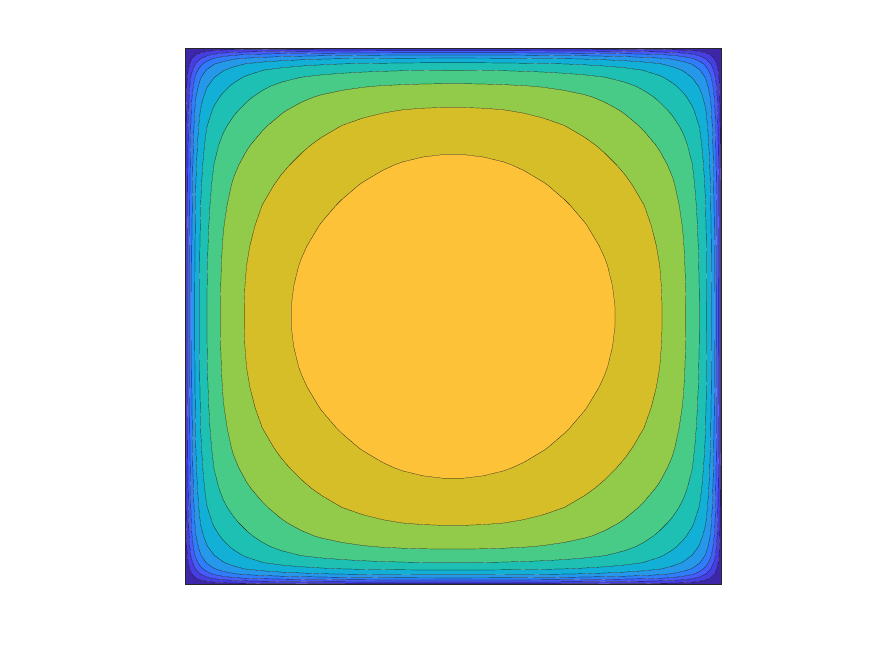}
    \caption{Baseline}
\end{subfigure}\,
\begin{subfigure}{0.45\textwidth}
    \includegraphics[width=\textwidth] {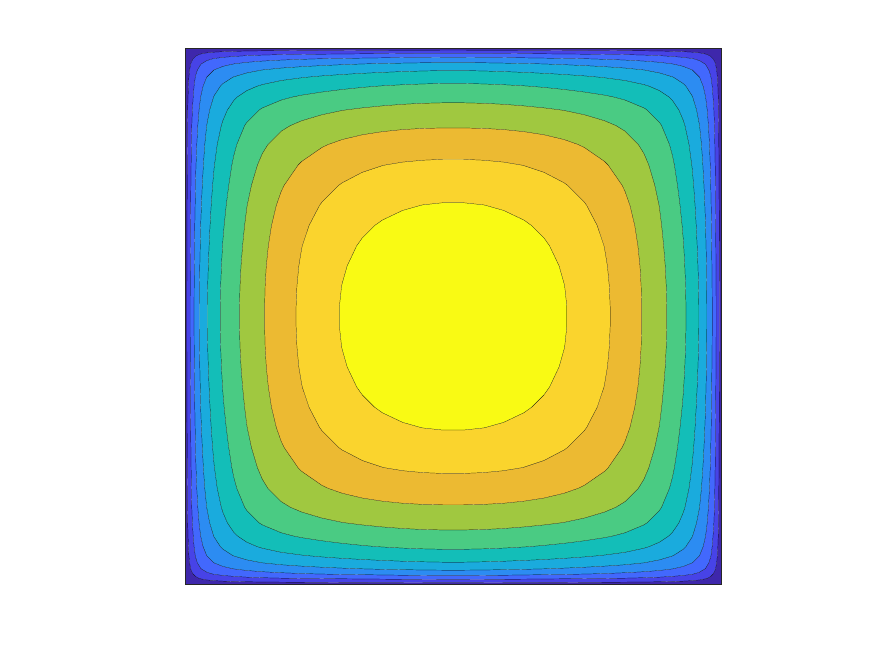}
    \caption{VBNN}
\end{subfigure}\,
\begin{subfigure}{0.45\textwidth}
    \includegraphics[width=\textwidth] {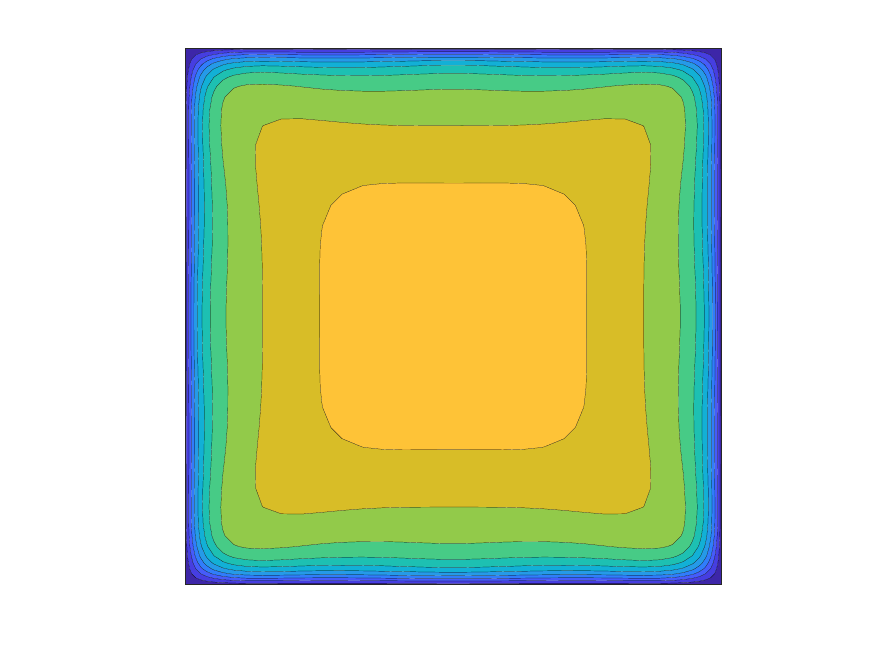}
    \caption{$\nu_t$-VBNN}
\end{subfigure}
\caption{Intensity of the streamwise velocity $u_x/U_b$.}
\label{Fig:Ux_section_nut_vbnn}
\end{figure}

The accurate description of the secondary motion affects the accuracy of the streamwise velocity, as depicted in Figure \ref{Fig:Ux_section_nut_vbnn}. As a matter of fact, when secondary motion is predicted, i.e. for all cases except the Baseline, the contour lines of $u_x/U_b$ have a square-like shape, while when secondary motion is not predicted such contour lines have a circle-like shape. In addition, this behavior is proportional to the intensity of the predicted secondary motion, making the contour lines of the $\nu_t$-VBNN case significantly closer to the DNS ones. This trend is physically expected: the secondary motion moves flow from the central region, characterized by higher $u_x/U_b$ values, toward the corners, thus increasing the streamwise velocity along the diagonals, making the contour lines square-shaped. Finally, we point out that the VBNN model overpredicts the intensity of the streamwise velocity in the cross-section center, while this is not the case for the $\nu_t$-VBNN model that agrees with the DNS results.

\begin{figure}[h]
\centering
\begin{subfigure}{0.325\textwidth}
    \includegraphics[width=\textwidth] {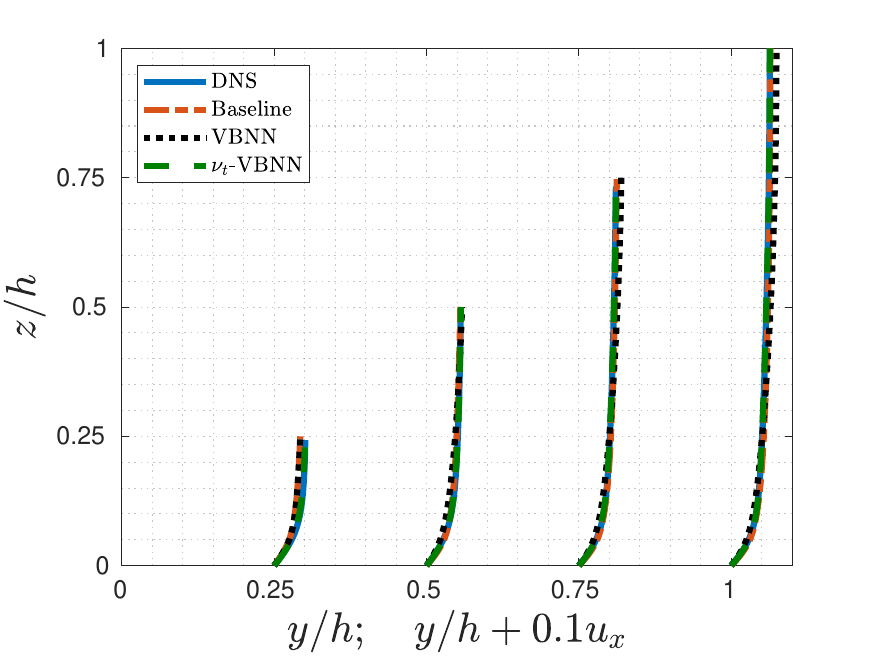}
    \caption{}
\end{subfigure}\,
\begin{subfigure}{0.325\textwidth}
    \includegraphics[width=\textwidth] {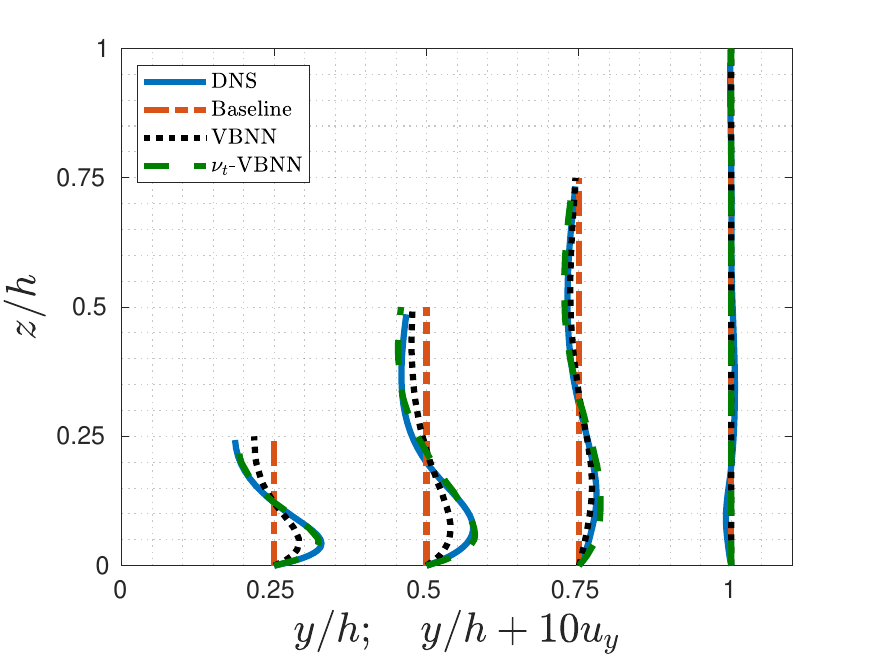}
    \caption{}
\end{subfigure}\,
\begin{subfigure}{0.325\textwidth}
    \includegraphics[width=\textwidth] {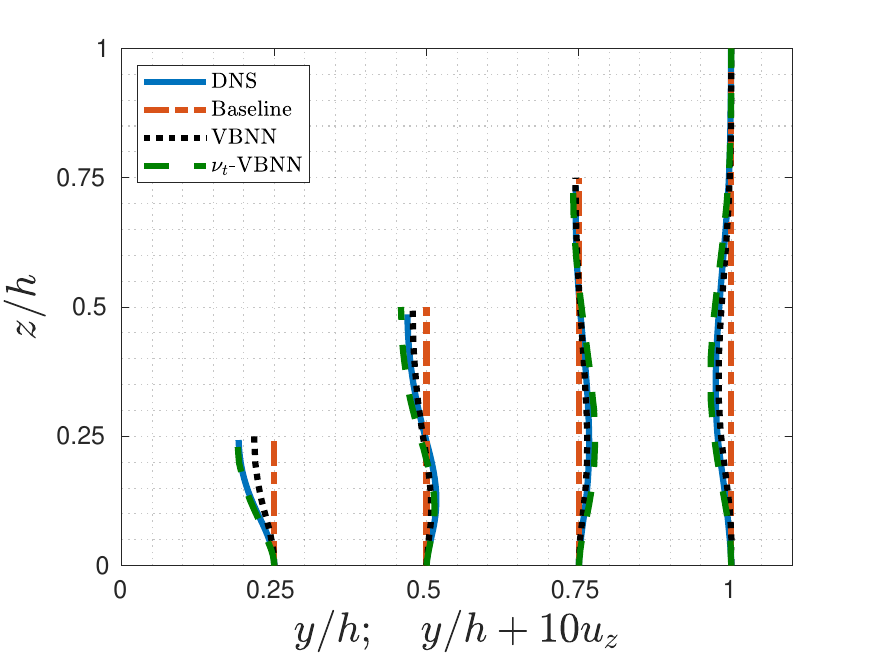}
    \caption{}
\end{subfigure}
\caption{Velocity profiles across the red lines in Figure \ref{fig:square_duct}.}
\label{fig:U_profiles}
\end{figure}

Figure \ref{fig:U_profiles} shows the profiles of the velocity components along the red lines over the square section in Figure \ref{fig:square_duct}. While the Baseline model correctly describe the $u_x$ velocity, it wrongly predicts negligible $u_y$ and $u_z$. This is not the case for the VBNN and $\nu_t$-VBNN approaches, being the latter consistently more accurate, particularly near the corner.

In general, we observed more accurate results when using the $\nu_t$-VBNN compared to the VBNN case. In the former, more attention is devoted during the training stage on the term that is treated implicitly in the momentum equation: a neural network is trained to predict this term only. In the latter, the term treated implicitly comes out from an optimization process involving other 11 quantities (all the coefficients of the linear combination \eqref{eq:lin_comb}). Because data-driven RANS models suffer of bad conditioning issues and the implicit treatment of the viscous contribution is the main solution to this issue, we think that focusing a training only for this term explains the improved accuracy of the $\nu_t$-VBNN model.

\section{Conclusions}
In this paper we present a new data-driven model called $\nu_t$-VBNN that is founded on the VBNN model presented in \cite{nostro}. This model predicts separately the turbulent viscosity $\nu_t$ and the Reynolds force vector component associated to the orthogonal part of the Reynolds stress tensor with respect to the mean Strain rate tensor $\orfv$. The $\nu_t$ field can be straightforwardly treated implicitly in the RANS equations improving the conditioning of the data-driven model. Both $\nu_t$ and $\orfv$ are predicted through feed forward neural networks that are trained using high-fidelity DNS data.

We compare the VBNN and $\nu_t$-VBNN models using the classic benchmark of flow in a square duct and we observe that the latter is significantly and consistently more accurate than the former.
 
\section*{Acknowledgements}
The author really thanks professor Stefano Berrone for his advices. The author is a member of the INdAM-GNCS research group CUP$\_$E53C23001670001.

\end{document}